\documentclass[a4paper]{jpconf}

\usepackage{hyperref,bm,amsmath,amssymb,a4wide,graphicx,subfigure,wasysym,amsfonts,cite}

\begin{document}

\title{Cosmological magnetic fields in turbulent matter}

\author{Maxim Dvornikov}

\address{Pushkov Institute of Terrestrial Magnetism, Ionosphere and Radiowave Propagation (IZMIRAN), 108840 Troitsk, Moscow, Russia}

\address{Physics Faculty, National Research Tomsk State University, 36 Lenin Avenue, 634050 Tomsk, Russia}

\ead{maxdvo@izmiran.ru}

\begin{abstract}
We study the evolution of magnetic fields in turbulent hot plasma of the early Universe accounting for the chiral magnetic effect. The magnetohydrodynamic  turbulence is modeled by replacing the matter velocity in the advection term in the Faraday equation with the Lorentz force. The system of the kinetic equations for the spectra of the densities of the magnetic helicity and the magnetic energy, as well as for the chiral imbalance, is derived. The amplification of the magnetic field is shown to result from the presence of the chiral magnetic effect solely. The system of the kinetic equations is solved numerically in primordial plasma after the electroweak phase transition. The influence of the matter turbulence on the magnetic field evolution is examined for different seed magnetic fields.
\end{abstract}

Magnetic fields are important for various physical processes including the cosmic rays propagation, the influence on the stellar and solar activities etc. Galactic magnetic fields, having the huge length scale comparable with a galaxy size, are supposed to be amplified from a seed magnetic field. The origin of this seed field is still an open problem in astrophysics and cosmology~\cite{Durrer:2013pga}. There is a possibility that seed magnetic fields are generated in the epoch of the structure formation. Another option, which is assumed in the present work, is that seed fields arise in the early Univese before the recombination.

The evolution of cosmic magnetic fields in primordial plasma, having the asymmetry of right and left particles, and the leptogenesis in (hyper-)magnetic fields before and after the electroweak phase transition (EWPT) were studied in~\cite{Boyarsky:2011uy,Dvornikov:2011ey,Dvornikov:2012rk}. The instability of the magnetic field in this primordial plasma, leading to the field generation, is owing to the chiral magnetic effect (CME)~\cite{Vilenkin}.

In this work, we shall discuss the influence of the matter turbulent motion on the generation of magnetic fields driven by the CME~\cite{DvoSem17}. The advection term in the Faraday equation was neglected in~\cite{Boyarsky:2011uy}. The MHD turbulence only was accounted for in~\cite{Campanelli:2007tc} while describing the evolution of magnetic fields in the early Universe. Recently the effect of the CME on the magnetic fields evolution in the turbulent primordial plasma was estimated numerically in~\cite{Bra17}.

We start with the system of the MHD equations in the single fluid approximation for the hot relativistic plasma in the early Universe after EWPT with the temperatures $10\,\mathrm{MeV}<T< T_\mathrm{EWPT}\simeq 100\,\mathrm{GeV}$. For plasma obeying the equation of state $P=\rho/3$, these equations have the form,
\begin{align}\label{MHD}
  & \partial_t\rho + \frac{4}{3}\nabla\cdot(\rho \mathbf{v})=0,
  \\
  & \frac{4}{3}\partial_t(\rho\mathbf{v}) -
  \frac{4}{3}\rho\mathbf{v}\times(\nabla\times \mathbf{v}) =
  (\mathbf{J}\times \mathbf{B}) - \nabla P +
  \frac{4}{3}\rho\nu\nabla^2\mathbf{v}, 
  \label{eq:NS}
  \\
  & \partial_t\mathbf{B} = -\nabla\times \mathbf{E},
  \label{eq:Bianchi} 
\end{align}
where $\rho$ is the energy density of the fluid, $P$ is the pressure, $\mathbf{B}$ is the magnetic field, $\mathbf{E}$ is the electric field, $\mathbf{v}$ is the fluid velocity, $\mathbf{J} = (\nabla \times \mathbf{B})$ is the electric current, and $\nu$ is the kinematic viscosity.

Since charged particles in plasma are ultrarelativistic, we have to account for the anomalous CME current $\mathbf{J}_\mathrm{CME} = (2\alpha_\mathrm{em}\mu_5/\pi)\mathbf{B}$ along with the usual ohmic current $\mathbf{J}_\mathrm{Ohm}=\sigma_\mathrm{cond}[\mathbf{E} + (\mathbf{v}\times \mathbf{B})]$ in equations~\eqref{MHD}-\eqref{eq:Bianchi}. Using the MHD approximation, we derive the Faraday equation modified by the CME contribution,
\begin{equation}\label{Faraday}
  \partial_t\mathbf{B} = \nabla \times (\mathbf{v}\times\mathbf{B}) +   
  \eta_m\nabla^2\mathbf{B} +
  \frac{2\alpha_\mathrm{em}\mu_5}{\pi \sigma_\mathrm{cond}}\nabla\times \mathbf{B}.
\end{equation}
Here $\alpha_\mathrm{em} \approx 7.3\times 10^{-3}$ is the fine structure constant, $\eta_m=\sigma_\mathrm{cond}^{-1}$ is the magnetic diffusion coefficient, $\sigma_\mathrm{cond}=\sigma_cT$ is the hot plasma conductivity with $\sigma_c\simeq 100$, $\mu_5 = (\mu_\mathrm{R} - \mu_\mathrm{L})/2$ is the chiral imbalance, and $\mu_\mathrm{R,L}$ are the chemical potentials of right and left electrons.

To proceed with the analysis of equation~\eqref{Faraday} we shall model the fluid velocity $\mathbf{v}$, obeying the Navier-Stokes equation~\eqref{eq:NS}, which is rather difficult to solve, with the Lorentz force $\mathbf{F}_\mathrm{L}\sim (\mathbf{J}\times \mathbf{B})$~\cite{DvoSem17},
\begin{equation}\label{velocity}
  \mathbf{v}=
  \frac{\tau_d}{P+ \rho}
  (\mathbf{J}\times \mathbf{B}),
\end{equation}
accompanied with the phenomenological drag time parameter $\tau_d \sim \alpha_\mathrm{em}^{-2}/T$~\cite{Cornwall:1997ms}. 

Basing on the master equations~(\ref{Faraday}) and~(\ref{velocity}), we derive the kinetic equations for the spectra of the magnetic energy density $\mathcal{E}_\mathrm{B}=\mathcal{E}_\mathrm{B}(k,t)$
and the density of the magnetic helicity $\mathcal{H}_\mathrm{B}=\mathcal{H}_\mathrm{B}(k,t)$~\cite{DvoSem17},
\begin{equation}\label{kinetics}
  \frac{\partial\mathcal{E}_\mathrm{B}}{\partial t} =
  -2k^{2}\eta_\mathrm{eff}\mathcal{E}_\mathrm{B}+\alpha_{+}k^{2}\mathcal{H}_\mathrm{B},
  \quad
  \frac{\partial\mathcal{H}_\mathrm{B}}{\partial t} =
  -2k^{2}\eta_\mathrm{eff}\mathcal{H}_\mathrm{B}+4\alpha_{-}\mathcal{E}_\mathrm{B},
\end{equation}
where
\begin{align}\label{parameters}
  \eta_\mathrm{eff} = &
  \sigma_\mathrm{cond}^{-1}+\frac{4}{3}\frac{\tau_{d}}{P+ \rho}
  \int \mathrm{d}p\mathcal{E}_\mathrm{B},
  \quad
  \alpha_{d} = \frac{2}{3}\frac{\tau_{d}}{P + \rho}
  \int \mathrm{d}p p^{2}\mathcal{H}_\mathrm{B},
  \nonumber
  \\
  \alpha_{\pm}  = &
  \alpha_\mathrm{CME} \mp \alpha_{d},
  \quad
  \alpha_\mathrm{CME} = \frac{\Pi(t)}{\sigma_\mathrm{cond}},
  \quad
  \Pi(t)=\frac{2\alpha_\mathrm{em}}{\pi}\mu_{5}(t).
\end{align}
The spectra $\mathcal{E}_\mathrm{B}$ and $\mathcal{H}_\mathrm{B}$ are related to the density of the magnetic energy and of the magnetic helicity by the following relations:
\begin{equation}\label{eq:EBHBdef}
  E_{\mathrm{B}}(t) = \frac{B^2}{2} = \int\mathrm{d}k\,\mathcal{E}_{\mathrm{B}}(k,t),
  \quad
  H_{\mathrm{B}}(t) = \frac{1}{V}\int \mathrm{d}^3 x ({\bf A} \cdot \mathbf{B}) =
  \int\mathrm{d}k\,\mathcal{H}_{\mathrm{B}}(k,t),
\end{equation}
where $V$ is the normalization volume and ${\bf A}$ is the vector potential. The details of the derivation of equations~\eqref{kinetics} and~\eqref{parameters} are provided in~\cite{DvoSem17}.

The difference of our results from findings of~\cite{Boyarsky:2011uy} is seen from the second non-linear terms in equation~(\ref{parameters}) which contain the drag time $\tau_{d}$ when we take into account the turbulent motion $\sim \mathbf{v}$. Note that the effective magnetic diffusion coefficient $\eta_\mathrm{eff}$ in equation~(\ref{parameters}) coincides with that in~\cite{Campanelli:2007tc} (accounting also for the factor $P+\rho$ in the denominator missed there). The analog of the $\alpha$-dynamo parameter, $\alpha_{\pm}$, in equation~(\ref{parameters}) differs from that derived in~\cite{Campanelli:2007tc} mainly because of the absence of the CME term there, and due to the different signs ($\pm$) in turbulent contributions for the evolution of the spectra $\mathcal{E}_\mathrm{B}$ and $\mathcal{H}_\mathrm{B}$ instead of the same sign ($+$) in both equations. 

When we study the evolution of magnetic fields in a hot plasma in the expanding universe, it is convenient to rewrite equations~\eqref{kinetics} and~\eqref{parameters} using the conformal dimensionless variables. They are introduced in the following way: $t\to \eta=M_0/T$ and $\tilde{k}=ak$, where $a=1/T$, $M_0=M_\mathrm{Pl}/1.66\sqrt{g^*}$, $M_\mathrm{Pl}=1.2\times 10^{19}\,\mathrm{GeV}$ is the Plank mass, and $g^{*}=106.75$ is the number of the relativistic degrees of
freedom. In these variables, equation~\eqref{kinetics}
takes the form,
\begin{equation}\label{eq:EBHBt}
  \frac{\partial\tilde{\mathcal{E}}_\mathrm{B}}{\partial\eta} =
  -2\tilde{k}^{2}\tilde{\eta}_\mathbf{eff}\tilde{\mathcal{E}}_\mathrm{B} +
  \tilde{\alpha}_{+}\tilde{k}^{2}\tilde{\mathcal{H}}_\mathrm{B},
  \quad
  \frac{\partial\tilde{\mathcal{H}}_\mathrm{B}}{\partial\eta} =
  - 2\tilde{k}^{2}\tilde{\eta}_\mathbf{eff}\tilde{\mathcal{H}}_\mathrm{B} +
  4\tilde{\alpha}_{-}\tilde{\mathcal{E}}_\mathrm{B},
\end{equation}
where $\tilde{\mathcal{E}}_\mathrm{B}=\tilde{\mathcal{E}}_\mathrm{B}(\tilde{k},\eta)$ and
$\tilde{\mathcal{H}}_\mathrm{B}=\tilde{\mathcal{H}}_\mathrm{B}(\tilde{k},\eta)$ are the conformal spectra, as well as $\tilde{\eta}_\mathrm{eff}$ and $\tilde{\alpha}_{\pm}$ are the effective magnetic diffusion and effective $\alpha$-dynamo parameters expressed in the conformal variables~\cite{DvoSem17}. The evolution equation for the chiral imbalance $\tilde{\mu}_{5}=\tilde{\mu}_{5}(\eta)$
has the form,
\begin{equation}\label{mu5}
  \frac{\mathrm{d}\tilde{\mu}_{5}}{\mathrm{d}\eta} =
  -\frac{6\alpha_\mathrm{em}}{\pi}
  \int \mathrm{d}\tilde{k}
  \frac{\partial\tilde{\mathcal{H}}_\mathrm{B}}{\partial\eta} -
  \tilde{\Gamma}_{f}\tilde{\mu}_{5},
  \quad
  \tilde{\Gamma}_{f} = 
  \alpha_\mathrm{em}^{2}\left(\frac{m_{e}}{3M_{0}}\right)^{2}\eta^{2},
\end{equation}
where take into account the helicity flip rate $\tilde{\Gamma}_{f}$ in a hot relativistic plasma~\cite{Boyarsky:2011uy}. 

We shall use the initial energy spectrum in the form, $\tilde{\mathcal{E}}_\mathrm{B}(\tilde{k},\eta_{0})=\mathcal{C}\tilde{k}^{\nu_\mathrm{B}}$.
The factor $\mathcal{C}$ can be found from the equation~\eqref{eq:EBHBdef}, rewritten in the conformal variables, by setting there $\tilde{B} = \tilde{B}_0$,
where $\tilde{B}_{0}=\tilde{B}(\eta=\eta_{0})$ is the seed magnetic
field. If we use the Batchelor initial spectrum with $\nu_\mathrm{B}=4$ and $0<\tilde{k}<\tilde{k}_\mathrm{max}=10^{-6}$, then
it is convenient to introduce the following dimensionless variables:
\begin{align}
  \mathcal{H}(\kappa,\tau) = &
  \frac{12\alpha_\mathrm{em}^{2}}{\pi^{2}}
  \tilde{\mathcal{H}}_\mathrm{B}(\tilde{k},\eta),
  \quad
  \mathcal{R}(\kappa,\tau) =
  \frac{24\alpha_\mathrm{em}^{2}}{\pi^{2}\tilde{k}_\mathrm{max}}
  \tilde{\mathcal{E}}_\mathrm{B}(\tilde{k},\eta),
  \quad
  \mathcal{M}(\tau) =
  \frac{2\alpha_\mathrm{em}}{\pi\tilde{k}_\mathrm{max}}\tilde{\mu}_{5}(\eta),
  \notag
  \\
  \tau = & \frac{2\tilde{k}_\mathrm{max}^{2}}{\sigma_{c}}\eta,
  \quad
  \kappa=\frac{\tilde{k}}{\tilde{k}_\mathrm{max}},
  \quad
  \mathcal{G} =
  \frac{\sigma_{c}}{2\tilde{k}_\mathrm{max}^{2}}\tilde{\Gamma}_{f}.
\end{align}
Using these variables, the system of kinetic equations takes the form,
\begin{align}
  \label{finalH}
  \frac{\partial\mathcal{H}}{\partial\tau}=  &  
  -\kappa^{2}\mathcal{H}
  \left[
    1+K_{d}\int_{0}^{1}\mathrm{d}\kappa'\mathcal{R}(\kappa',\tau)
  \right]+
  \mathcal{R}
  \left[
    \mathcal{M}+
    K_{d}\int_{0}^{1}\mathrm{d}\kappa'\kappa^{\prime2}\mathcal{H}(\kappa',\tau)
  \right],
  \\
  \label{finalR}
  \frac{\partial\mathcal{R}}{\partial\tau}= &
  -\kappa^2\mathcal{R}
  \left[
    1+K_{d}\int_{0}^{1}\mathrm{d}\kappa'\mathcal{R}(\kappa',\tau)
  \right]+
  \kappa^{2}\mathcal{H}
  \left[
    \mathcal{M}-
    K_{d}\int_{0}^{1}\mathrm{d}\kappa'\kappa^{\prime2}\mathcal{H}(\kappa',\tau)
  \right],
  \\
  \label{finalM}
  \frac{\mathrm{d}\mathcal{M}}{\mathrm{d}\tau} = & 
  \int_{0}^{1}\mathrm{d}\kappa
  \left(
    \kappa^{2}\mathcal{H}-\mathcal{R}\mathcal{M}
  \right) -
  \mathcal{G}\mathcal{M},
\end{align}
where $K_{d}=5\sigma_{c}\tilde{k}_\mathrm{max}^{2}/4\alpha_\mathrm{em}^{4}g^{*}$. It is interesting to note that the contribution of the turbulent terms cancels out in equation~\eqref{finalM}. Nevertheless there is a turbulence contribution in equation~\eqref{finalH}. The initial values of the functions $\mathcal{R}$ and $\mathcal{H}$
are $\mathcal{R}(\kappa,\tau_{0})=\mathcal{R}_{0}\kappa^{\nu_\mathrm{B}}$
and $\mathcal{H}(\kappa,\tau_{0})=q\mathcal{R}_{0}\kappa^{\nu_\mathrm{B}-1}$,
where $\mathcal{R}_{0} = 12\alpha_\mathrm{em}^{2}\tilde{B}_{0}^{2}(\nu_\mathrm{B}+1)/\pi^{2}\tilde{k}_\mathrm{max}^{2}$ and the parameter $0\leq q\leq1$ fixes the initial magnetic helicity.

\begin{figure}
  \centering
  \subfigure[]
  {\label{1a}
  \includegraphics[scale=.11]{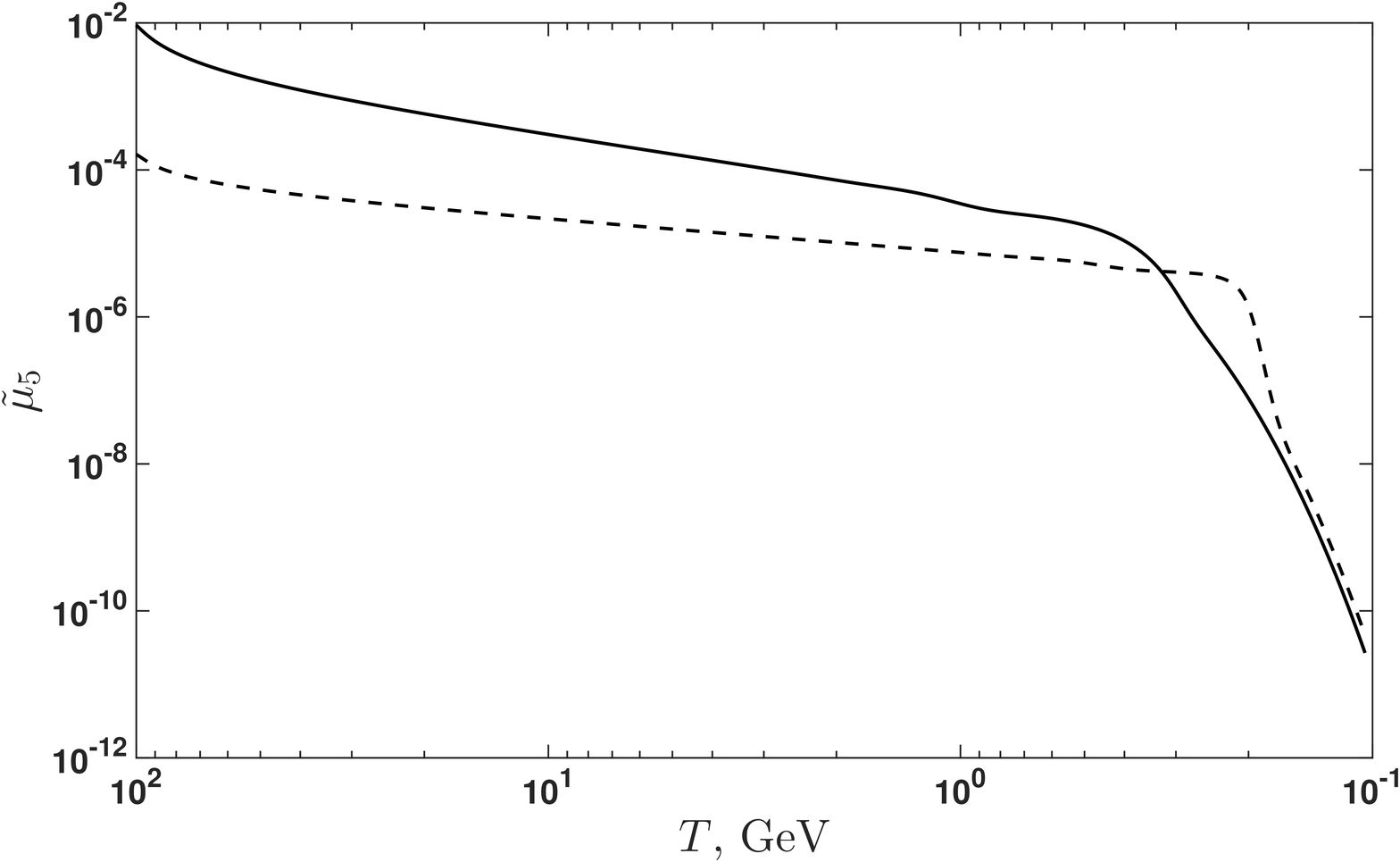}}
  \hskip-.7cm
  \subfigure[]
  {\label{1b}
  \includegraphics[scale=.11]{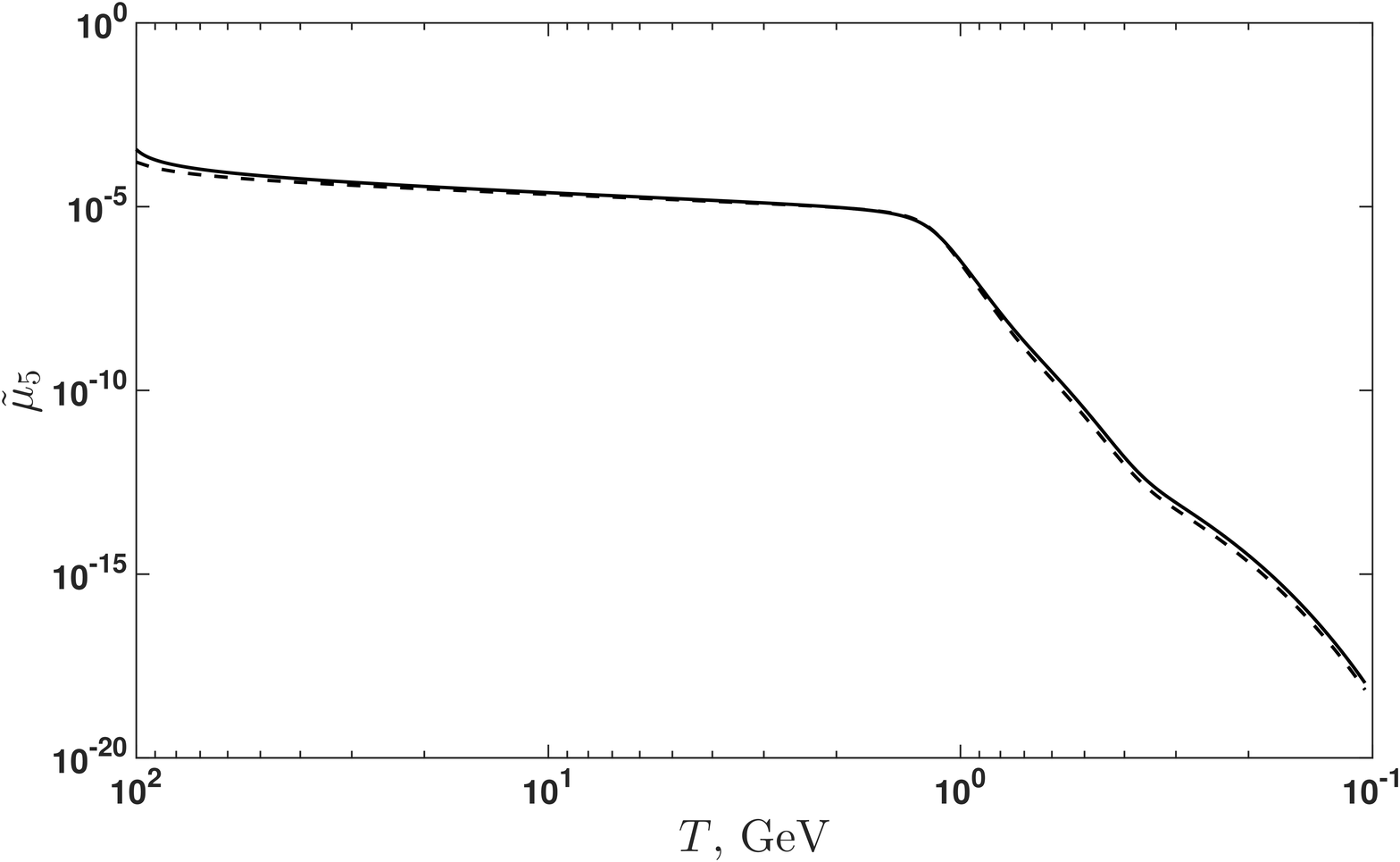}}
  \\
  \subfigure[]
  {\label{1c}
  \includegraphics[scale=.11]{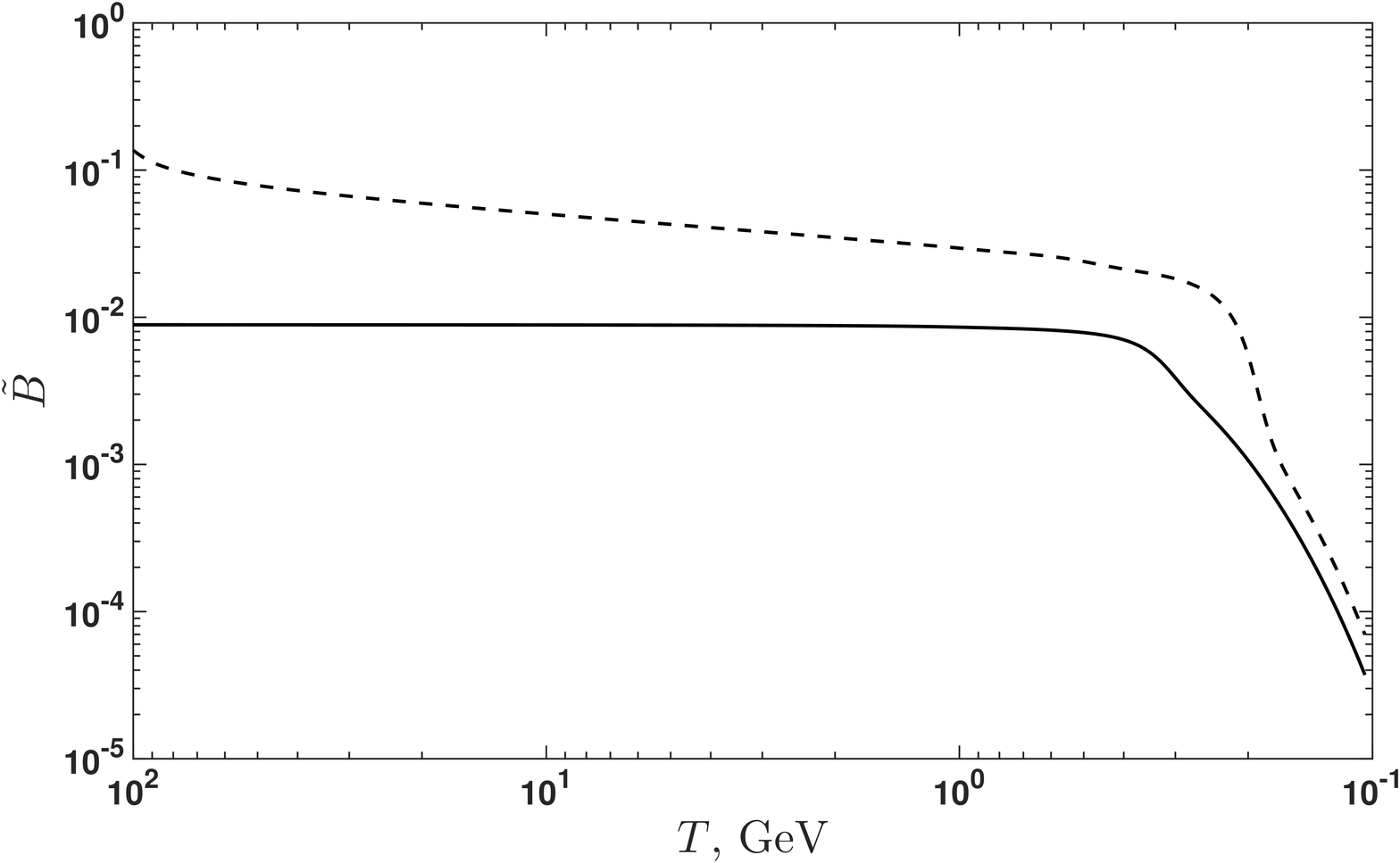}}
  \hskip-.7cm
  \subfigure[]
  {\label{1d}
  \includegraphics[scale=.11]{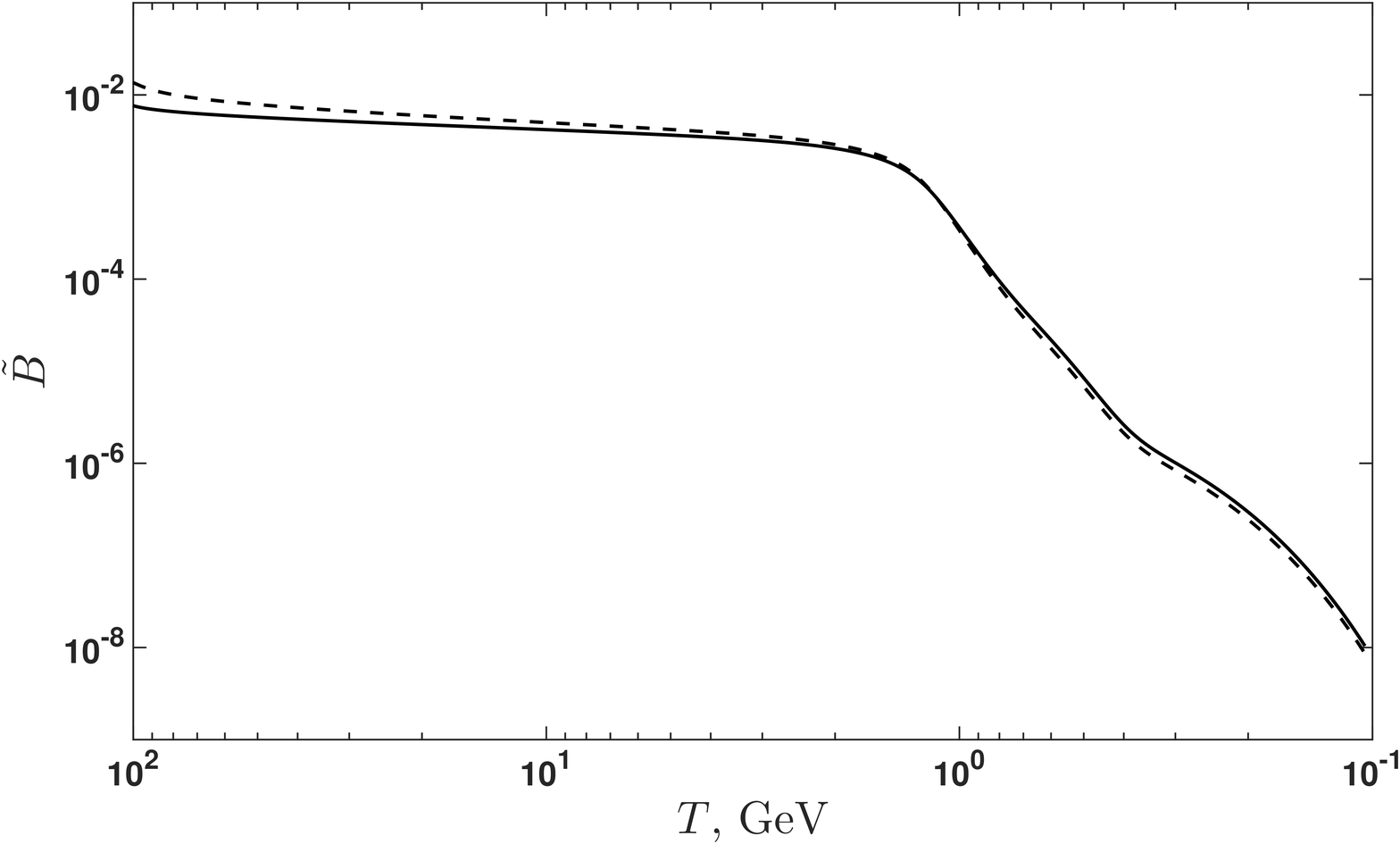}}
  \\
  \subfigure[]
  {\label{1e}
  \includegraphics[scale=.11]{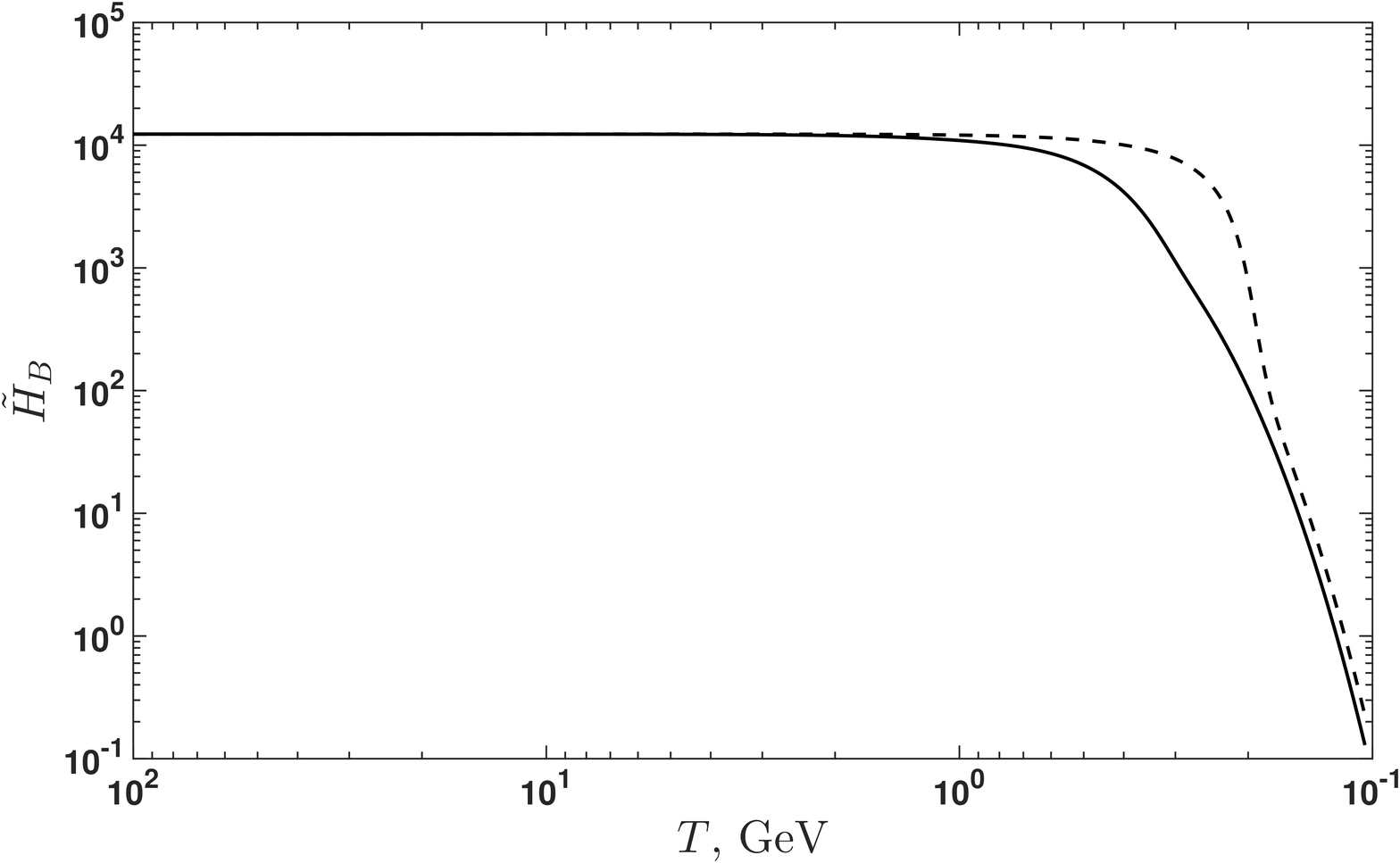}}
  \hskip-.7cm
  \subfigure[]
  {\label{1f}
  \includegraphics[scale=.11]{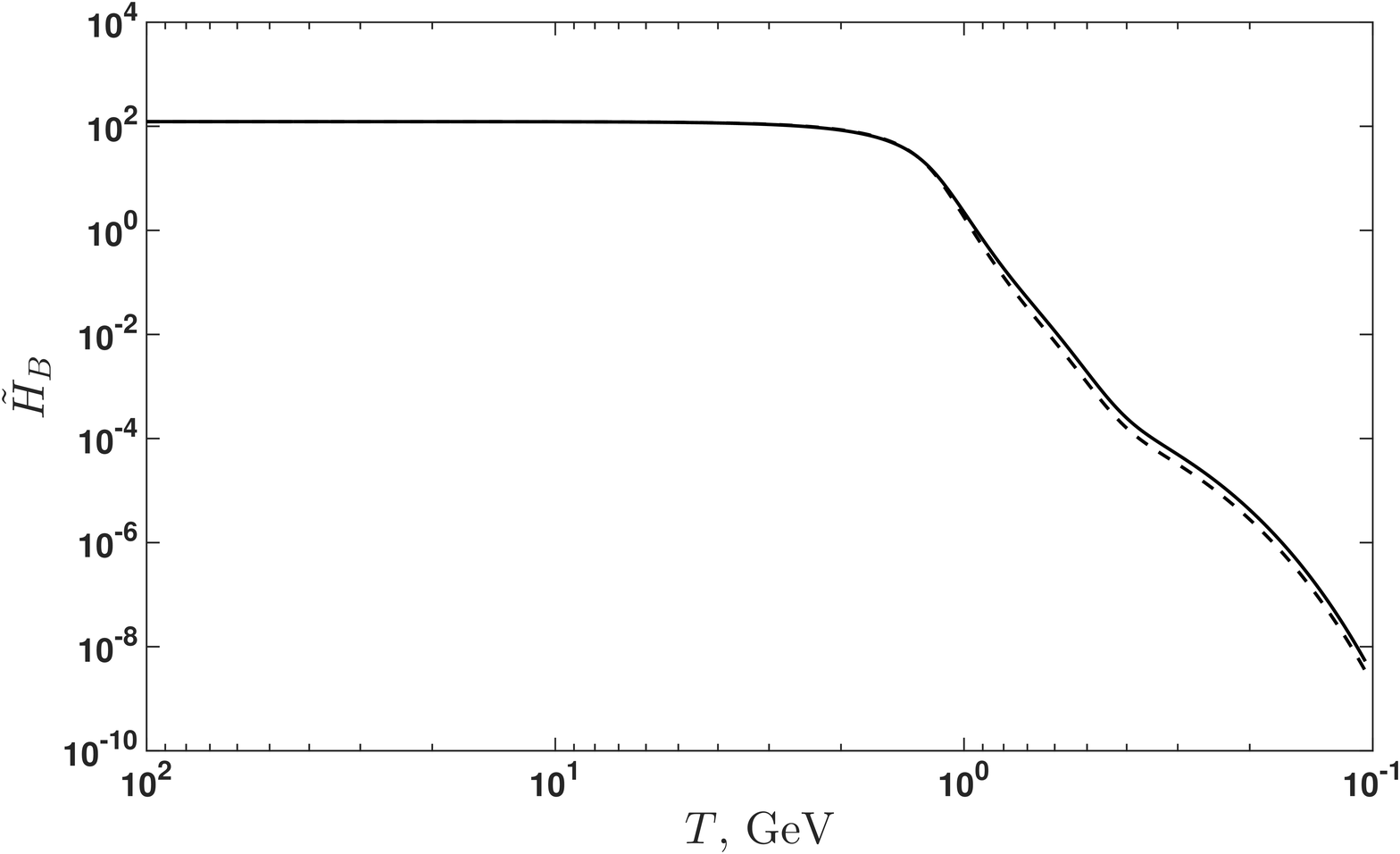}}
  \protect
  \caption{\label{fig:Bevol}
  The evolution of the chiral imbalance, the magnetic energy density
  and the helicity density in plasma of the early universe
  at $10^2\,\mathrm{MeV} < T < 10^2\,\mathrm{GeV}$.
  (a) and (b) The evolution of the chiral imbalance.
  (c) and (d)  The evolution of the magnetic field.
  (e) and (f)  The evolution  magnetic helicity density.
  Panels (a), (c), and (e) correspond to $\tilde{B}_0=10^{-1}$ whereas
  panels (b), (d), and (f) to $\tilde{B}_0=10^{-2}$.
  Solid lines show the evolution accounting for both
  the turbulence effects ($K_d \neq 0$) and the CME, whereas dashed lines
  are for the CME case only ($K_d = 0$).}
\end{figure}

We solve kinetic equations~(\ref{finalH})-\eqref{finalM} numerically.
The influence of the turbulent matter motion $\sim \mathbf{v}$ on the MHD characteristics, such as the magnetic field strength and the magnetic helicity, as well as on the chiral asymmetry parameter $\mu_5(t)$ in a hot plasma in the broken phase of early Universe is illustrated in figure~\ref{fig:Bevol}. The solid lines correspond to the case where both effects, i.e. the CME and the turbulent motion of matter $\mathbf{v}\sim \tau_d$, are taken into account, while the dash lines correspond to the CME effect only applied in~\cite{Boyarsky:2011uy}. Note that we present the numerical solutions of equations~(\ref{finalH})-\eqref{finalM} for the maximum helicity parameter $q=1$ only. In figures~\ref{1a}, \ref{1c} and~\ref{1e} we show results for the maximum initial magnetic field $\tilde{B}_0=0.1$ still obeying the BBN bound on the magnetic field~\cite{Cheng:1993kz}. In figures~\ref{1c}, \ref{1d} and~\ref{1f} we show the results for a smaller seed field $\tilde{B}_0=10^{-2}$.

One can see in figure~\ref{fig:Bevol} that the stronger the initial magnetic field is, the more noticeable the difference between the turbulent and nonturbulent cases is. 
The dependence of $\mu_5$ on the turbulence parameter $K_d$ is hidden non-trivially 
in equation~\eqref{finalM}. Such a hidden dependence comes rather from the magnetic field characteristics $\mathcal{H}$ and $\mathcal{R}$, which evidently depend on that parameter as seen in equations~(\ref{finalH}) and~\eqref{finalR}. While the diffusion terms for the magnetic helicity $\mathcal{H}$ and the magnetic energy densities spectra $\mathcal{R}$ are both enhanced  by turbulent motions $\sim K_d$, the instability (generation) terms $\sim \mathcal{M}\sim \tilde{\mu}_5$ are supplemented differently through the same parameter $K_d$. The magnetic helicity $\mathcal{H}$ is supported by turbulent motions even for a decreasing chiral anomaly $\mu_5$; cf. figures~\ref{1e} and~\ref{1f}. The magnetic energy $\mathcal{R}$ reduces additionally through the turbulent parameter $K_d$. This is a reason why the solid curve for magnetic field strength in figures~\ref{1c} and~\ref{1d} occurs below the dash curves corresponding to the pure CME effect.

Let us stress that such opposite contributions of the turbulent motion $\sim \mathbf{v}$ to the evolution of $\mathcal{H}$ and $\mathcal{R}$ comes directly from different signs of the parameter $\alpha_d$ in equation~(\ref{parameters}) as we found in contrast to the results in~\cite{Campanelli:2007tc}. Another important result obtained in~\cite{DvoSem17} is the examination of the possibility for the plasma turbulence to drive the magnetic field instability. In our work we have approximated the plasma velocity by the Lorentz force. We can see that, if one accounts for only the plasma turbulence contribution, i.e. assuming that $\alpha_d \neq 0$ and $\alpha_\mathrm{CME} = 0$, the initial magnetic field cannot be amplified. This our new finding confronts the statement of~\cite{Campanelli:2007tc}, where it was claimed that plasma turbulence described within the chosen model can provide the enhancement of a seed magnetic field.

\ack

I am thankful to the Tomsk State University Competitiveness Improvement Program and RFBR
(research project No.~15-02-00293) for a partial support.

\end{document}